\documentstyle[12pt]{article}

\def\thefootnote{\fnsymbol{footnote}}

\newlength{\minitwocolumn}
\catcode`\@=11
\long\def\@makefntext#1{
\protect\noindent \hbox to 3.2pt {\hskip-.9pt  
$^{{\eightrm\@thefnmark}}$\hfil}#1\hfill}               

\def\thefootnote{\fnsymbol{footnote}}
\def\@makefnmark{\hbox to 0pt{$^{\@thefnmark}$\hss}}    
        
\def\ps@myheadings{\let\@mkboth\@gobbletwo
\def\@oddhead{\hbox{}
\rightmark\hfil\eightrm\thepage}   
\def\@oddfoot{}\def\@evenhead{\eightrm\thepage\hfil
\leftmark\hbox{}}\def\@evenfoot{}
\def\sectionmark##1{}\def\subsectionmark##1{}}

\font\eightrm=cmr8

\def\brs{\delta}

\def\dM{{d_M}}

\newcommand{\rd}{\overleftarrow{\partial}} 
\newcommand{\ld}{\overrightarrow{\partial}} 

\newcommand{\sbv}[2]{{({{#1},{#2}})}}

\def\ba{\mbox{\boldmath $A$}}
\def\bb{\mbox{\boldmath $B$}}
\def\by{\mbox{\boldmath $Y$}}
\def\bz{\mbox{\boldmath $Z$}}
\def\bR{\mbox{\boldmath $R$}}

\def\bphi{\mbox{\boldmath $\phi$}}

\newcommand{\Q}{{\kern.24em\vrule width.04em height1.4ex%
                 depth-.05ex\kern-.26em\mathsf Q}}
\newcommand{\C}{{\kern.24em\vrule width.04em height1.4ex%
                 depth-.05ex\kern-.26em\mathsf C}}

\begin{document}


\baselineskip 0.7cm

\begin{titlepage}
\begin{flushright}
\end{flushright}

\vskip 1.35cm
\begin{center}
{\Large \bf
Topological Membranes with $3$-Form $H$ Flux \\
on Generalized Geometries
}
\vskip 1.2cm
Noriaki IKEDA$^1$%
\footnote{E-mail address:\ ikeda@yukawa.kyoto-u.ac.jp}
\ and Tatsuya TOKUNAGA$^2$
\footnote{E-mail address:\ tokunaga@yukawa.kyoto-u.ac.jp}
\vskip 0.4cm
{\it $^1$Department of Mathematical Sciences,
Ritsumeikan University \\
Kusatsu, Shiga 525-8577, Japan }\\
{\it $^2$Yukawa Institute for Theoretical Physics,
Kyoto University \\
Kyoto 606-8502, Japan }\\
\date{}

\vskip 1.5cm

\begin{abstract}
We construct topological string and topological membrane actions 
with a nontrivial $3$-form flux $H$ 
in arbitrary dimensions.
These models realize Bianchi identities with a nontrivial 
$H$ flux as consistency conditions.
Especially, we discuss the models with 
a generalized $SU(3)$ structure,
a generalized $G_2$ structure and 
a generalized $Spin(7)$ structure.
These models are constructed from the AKSZ formulation of 
Batalin-Vilkovisky formalism.
\end{abstract}
\end{center}
\end{titlepage}

\renewcommand{\thefootnote}{\alph{footnote}}

\setcounter{page}{2}


\rm

\section{Introduction}
\noindent
Recently, flux vacua in superstring theory have been studied.
Some superstring background geometries with a nonzero NS-NS $3$-form flux $H = \dM b \neq 0$ are called 
generalized geometries, which were proposed by Hitchin as a generalization of Calabi-Yau geometry\cite{Hit}.
%
%
%
%
Also, $N=(2, 2)$ supersymmetric sigma model with a nontrivial NS 
$B$-field has a bi-Hermitian structure\cite{GHR}, which was proved to be equivalent to 
a generalized K\"ahler structure \cite{Gua}.
%
By use of topological twisting, topological sigma model
with a generalized Calabi-Yau structure was constructed\cite{Kap}\cite{KL}\cite{Chuang:2006vt}\cite{Zucchini:2006ii}.
Other types of topological sigma models with a generalized complex structure 
were proposed and analyzed in
\cite{LMTZ}\cite{Zucchini:2004ta}\cite{Bergamin:2004sk}\cite{Zucchini:2005rh}\cite{Lindstrom:2005zr}\cite{Bredthauer:2006hf}\cite{Pestun:2006rj}\cite{Guttenberg:2006zi}.
Moreover, topological M theory with a $G_2$ structure on a $7$ manifold
\cite{Dijkgraaf:2004te}\cite{Anguelova:2005cv}\cite{Bao:2005yx}\cite{Nekrasov:2005bb} and topological F theory with a $Spin(7)$ structure on a $8$ manifold \cite{Anguelova:2004wy} were studied. 
%
In \cite{Alekseev:2004np}\cite{Bonelli:2005ti}\cite{Bonelli:2005rw}, 
by studying current algebra, topological membrane models with $SU(3)$, $G_2$ and $Spin(7)$ structures were studied.   
%

 As a next step after these good works, we would need to construct topological string and topological membrane theory with $H \neq 0$.  
In this paper, we construct topological membrane actions on generalized 
geometries, including topological $M$ and topological $F$ theory with $H$-flux.  Then we use the definitions of generalized exceptional structures, which were introduced 
by Witt \cite{Witt:2004vr}\cite{Witt:2005sk} as generalizations of 
$G_2$ and $Spin(7)$ structures.  Here generalized $G_2$ and $Spin(7)$ structures mean to add a $B$ field.  These generalized exceptional structures appear in 
background geometries of flux compactifications in type II superstring and M theory \cite{Jeschek:2004wy}\cite{Jeschek:2005ek}\cite{Tsimpis:2005kj}.
In this paper, for examples, we show that generalized $SU(3)$, $G2$ and $Spin(7)$ structures appear from consistency conditions of our actions.  
Moreover, dimensional reductions of these models are studied.  For example, generalized topological $G2$ models can be related to two kinds of generalized topological $SU(3)$ models, which are called A-type and B-type.      
Also, Batalin-Vilkovisky formalism\cite{Batalin:1981jr}\cite{Batalin:1984jr}
is one of the most general and the most theoretical settings to construct a field theory.
Especially, some topological field theories can 
be systematically formulated by 
the AKSZ formulation\cite{Alexandrov:1995kv}
of Batalin-Vilkovisky formalism.  
In the last part of this paper, 
we rewrite our actions by using superfield AKSZ formulation
of Batalin-Vilkovisky formalism.  
Then we use an extension of 
the Batalin-Vilkovisky construction of topological theories
to in any dimensions \cite{Ikeda:2001fq}\cite{Ikeda:2006wd}.

This paper is organized as follows.
In section 2, we briefly explain generalized geometry, which means to have a $H$ flux.  
The definition of generalized structures (\ref{ggequation}) is similar to Bianchi identities for RR fluxes in type II superstring background with a NS-NS flux.  
(See \cite{Grana:2005jc} as a review.) 
In section 3, we propose topological sigma models
with generalized structures (\ref{ggequation})
on arbitrary dimensional target spaces $M$.
In section 4, as some examples, we show that these models have generalized $SU(3)$ structures on $6$ manifolds, 
generalized $G_2$ structures on $7$ manifolds 
and generalized $Spin(7)$ structures on $8$ manifolds.
In section 5, we discuss two kinds of dimensional reductions.  
Generally, we derive lower dimensional generalized structures from higher dimensional ones.  
In section 6, we propose an extension of the AKSZ construction of Batalin-Vilkovisky formalism.
In section 7, we rewrite topological sigma models with generalized structures by use of superfield 
Batalin-Vilkovisky formalism.

\section{Geometries with $3$-Form Flux}
\noindent
In this section, we briefly summarize geometries
with $H$ flux.

We define a geometry with $H$ flux, which is a closed $3$-form.  Let $M$ be a manifold in $d$ dimensions with 
local coordinates $\{ \phi^i \}$. 
Let $\Lambda (M) = \Lambda^* T^* M$ be the space of 
all exterior differential forms on $M$.
%
%
%
%
If $\rho \in \Lambda(M)$ on $M$ satisfies following equations
\begin{eqnarray}
&&\dM H = 0,  
\label{Hequation}
\\
&&
(\dM + H) \wedge \rho = 0,
\label{ggequation}
\end{eqnarray}
where $\dM$ is an exterior differentiation on $M$, we call 
a {\it generalized 
(geometry) structure}.  
$\rho$ is expanded as formal sum of $p$-forms as
\begin{eqnarray}
\rho = \sum_{p=0}^{d} \rho_p, 
\label{rhoexpansion}
\end{eqnarray}
where $\rho_p$ is a $p$-form on a target space $M$,
i.e.
$\rho_p = \frac{1}{p!} \rho_{p, i_1 \cdots i_p}(\phi)
d \phi^{i_1} \cdots d \phi^{i_p}$.
%
After substituting  (\ref{rhoexpansion}) for (\ref{ggequation}), 
the equation (\ref{ggequation}) is separated to two independent sets of equations.  
The equations for even $p$-forms are 
\begin{eqnarray}
\dM \rho_0 = 0, \qquad
\dM \rho_2 + H \wedge \rho_0 = 0, \qquad
\dM \rho_4 + H \wedge \rho_2 = 0, \qquad
\cdots
\label{Atype}
\end{eqnarray}
The equations for odd $p$-forms are
\begin{eqnarray}
\dM \rho_1 = 0, \qquad
\dM \rho_3 + H \wedge \rho_1 = 0, \qquad
\dM \rho_5 + H \wedge \rho_3 = 0, \qquad
\cdots
\label{Btype}
\end{eqnarray}
Here we call the equations (\ref{Atype}) {\it A-type},
and the equations (\ref{Btype}) {\it B-type}.  
Also, $\Lambda (M)$ can be decomposed to irreducible representation 
spaces $\Lambda^{ev} (M)$ and $\Lambda^{od} (M)$ for spinor representation.  
$\Lambda^{ev} (M)$ is the sum of representation spaces of even forms.  
$\Lambda^{od} (M)$ is the sum of representation spaces of odd forms.  
Here we call $\rho \in \Lambda^{ev} (M)$ which satisfies 
(\ref{ggequation}) as {\it A-type} or even type, 
and $\rho \in \Lambda^{od} (M)$ which satisfies 
(\ref{ggequation}) as {\it B-type} or odd type.  
%
%
%
%
%
%
%
%
%
%
%
Especially, on a 6 manifold $M$,
we call ($\rho \in \Lambda^{ev} (M)$ 
a A-type generalized $SU(3)$ structure, 
and $\rho \in \Lambda^{od} (M)$ 
a B-type generalized $SU(3)$ structure.  
On a 7 manifold $M$, 
we call $\rho \in \Lambda^{ev} (M)$ 
a A-type generalized $G_2$ structure, 
and $\rho \in \Lambda^{od} (M)$ 
a B-type generalized $G_2$ structure.  
On a 8 manifold $M$, 
we call $\rho \in \Lambda^{ev} (M)$ 
a A-type generalized $Spin(7)$ structure, 
and $\rho \in \Lambda^{od} (M)$ 
a B-type generalized $Spin(7)$ structure.  
We will see that these examples of generalized structures appear from our models.



\section{Topological Membranes with $3$-Form Flux}
\noindent
In this section, we propose membrane action whose target space 
has a generalized structure.

We consider 
a smooth map $\{\phi^i \}$ from a $n$ dimensional worldvolume $\Sigma_n$ to a target manifold $M$
in $d$ dimensions.  
Let $\rho$ be a formal sum of forms on $M$.  
Then $\rho$ can be expanded for the form degrees as 
$\rho = \sum_{p=0}^{d} \rho_p$,
where $\rho_p$ is a $p$-form on $M$.
By use of local coordinates $\rho_p$ can be represented as 
$\rho_p = \frac{1}{p!} \rho_{p, i_1 \cdots i_p}(\phi)
d \phi^{i_1} \cdots d \phi^{i_p}$,
where $d$ is a exterior differentiation on $\Sigma_n$.  Also, let $H = H_{ijk} d \phi^i d \phi^j d \phi^k$ 
be a NS $3$-form which satisfies $d_M H = 0$.
~From closeness of $H$, $H$ can locally be expressed by 
a $2$-form B-field $b$ as $H = \frac{1}{2} d_M b$,
where
$b = b_{ij} d \phi^i d \phi^j$.  
%
%

Here, in order to construct
sigma models we introduce an auxiliary field $Y_2$, which is a $2$-form on $\Sigma_n$ and a scalar on $M$,
i.e. $Y_2 \in \wedge^2 T^* \Sigma_n$.  
Then we propose following action:
\begin{eqnarray}
S = \int_{\Sigma_n}
e^{Y_2} e^{\frac{1}{2} b} \rho
= \sum_{p=0}^{d} 
\frac{1}{p!} \int_{\Sigma_n} \  
e^{Y_2} e^{\frac{1}{2} b} \  
\rho_{p, i_1 \cdots i_p}(\phi)
d \phi^{i_1} \cdots d \phi^{i_p}.
\label{ggaction}
\end{eqnarray}
Note that enough higher orders of the exponentials vanish 
because $\Sigma_n$ has a finite number of dimensions.  
%
Also, $Y_2$ and $b$ are even forms, so if $n$ is even or odd, only even or odd parts of  $\rho_{p}$ remain.  
This means that the part $\int_{\Sigma_n} e^{Y_2}$ projects $\rho$ ($\in \Lambda (M)$ )
to $\Lambda^{od} (M)$ or $\Lambda^{ev} (M)$.  
This action can be topological if enough gauge symmetry exists to exclude all degree freedom of $\phi^i$.

Next we consider this condition that gauge symmetry remains. 
First, we define canonical conjugate momenta for $\phi^i$ and $Y_2$:
\begin{eqnarray}
&& \pi_{\phi i} = 
e^{Y_2} e^{\frac{1}{2} b} 
\left(
b_{ij} d \phi^j \rho +
\sum_{p=1}^{d} 
\frac{1}{(p-1)!} \rho_{p, i j_2 \cdots j_p}(\phi)
d \phi^{j_2} \cdots d \phi^{j_p} 
\right), 
\nonumber \\
&& 
\pi_{Y_2} = 0.
\end{eqnarray}
We defines gauge symmetry generators $G_i$:
\begin{eqnarray}
&& G_i = \pi_{\phi i} - 
e^{Y_2} e^{\frac{1}{2} b} 
\left(
b_{ij} d \phi^j \rho +
\sum_{p=1}^{d} 
\frac{1}{(p-1)!} \rho_{p, i j_2 \cdots j_p}(\phi)
d \phi^{j_2} \cdots d \phi^{j_p} 
\right),
\nonumber \\
&& 
G_{Y_2} = \pi_{Y_2}.
\end{eqnarray}
Then this system has the constraints $G_i \approx 0$
and $G_{Y_2} \approx 0$ on the phase space 
$(\phi^i, Y_2, \pi_{\phi i}, \pi_{Y_2})$.  
We require that these constraints should be the first class 
in order to make a consistent theory, i.e.
$
\{ G_I, G_J \}  \approx 0
$,
where $\{ \cdot, \cdot \}$ is the Poisson bracket 
on the phase space.   Here $G_I$ is $(G_i,G_{Y_2})$, and 
$\approx$ means that the equation is satisfied on 
the constraint surface.  
This condition is
\begin{eqnarray}
&& 
\frac{\partial \rho_{p+2, i_1 \cdots i_{p+2}}}{\partial \phi^j} (\phi)
+ \frac{3!}{(p+2)(p+1)} 
H_{[ji_{p+1}i_{p+2}}(\phi) \rho_{p, i_1 \cdots i_{p}]}(\phi)
= 0.
\label{generalizedintegrablecomponent}
\end{eqnarray}
Note that this condition is just equal to the generalized structure condition (\ref{ggequation}).  
The gauge transformation is derived from the 
Poisson bracket with $G_i$.
The number of $G_i$ is equal to the number of $\phi^i$.  If the gauge symmetry exists, this model is topological. 
Therefore, from the condition that this action is topological, we can get the generalized structure condition on the target geometry with $3$-form $H$ flux.  
We can construct topological membrane theory on generalized geometry.  
%
%
%

Moreover, we can rewrite this action using a $1$-st order formalism.  
We introduce four auxiliary fields.  
The first one is a $1$-form $A_1{}^{i}$ such that $A_1{}^{i} = d \phi{}^i$.  
The second is a $n-1$-form field $B_{n-1,i}$ as a Lagrange multiplier field.  
%
%
The others are an $n-2$-form $B_{n-2,i}$ and an $n-3$-form $Z_{n-3}$, which are introduced in order to make kinetic terms for $A_1{}^{i}$ and $Y_2$, like $(-1)^{n-1} B_{n-2,i} d A_1{}^i - Y_{2} d Z_{n-3}$.  
These kinetic terms are consistent with the equations of motion, and do not change constraints (\ref{generalizedintegrablecomponent}).  
Then we can write following $1$-first order action:
\begin{eqnarray}
S &=& \int_{\Sigma_n}
(-1)^{n} B_{n-1,i} d \phi{}^i + (-1)^{n-1} B_{n-2,i} d A_1{}^i
- Y_{2} d Z_{n-3}
+ (-1)^{n-1} B_{n-1,i} A_1{}^{i}
\nonumber \\
&& 
+ \exp{(Y_2)} \exp{\frac{1}{2} (b_{ij} A_1{}^i A_1{}^j)}
\biggl(
\sum_{p=0}^{d}
\frac{1}{p!} \rho_{p, i_1 \cdots i_p}(\phi)
A_1{}^{i_1} \cdots A_1{}^{i_p}
\biggr),
\label{ndaction}
\end{eqnarray}
where $(-1)^{n}$ is a convention for sign factor.  
We will analyze this first order action by use of the Batalin-Vilkovisky(BV) formalism in the latter part.

\section{Topological String, Topological $M$ and Topological $F$ Theories on Generalized geometries}
\noindent
In this section, as examples of out models we consider the cases that the target spaces have 6, 7 and 8 dimensions.  
We call these topological models as topological string (membrane), topological M or topological F theories, respectively.  
\subsection{Topological Strings and Membranes on 
Generalized $SU(3)$-Manifolds}
\noindent
First as the first example, we consider the case of topological membrane theory with a generalized $SU(3)$ structure 
on a $6$ dimensional target space $M$.  
Let $M$ be a six-dimensional manifold.  

We shortly review $A$ model and $B$ model in the case of $H=0$.  $A$-model
is constructed from a K\"ahler structure $\rho_{2}$ on $M$.
The $A$ model action is 
\begin{eqnarray}
S = \int_{\Sigma_2}
\frac{1}{2!} \rho_{2, ij}
d \phi^{i} d \phi^{j}.
\label{su3action}
\end{eqnarray}
The first class constrain is $\dM \rho_{2} = 0$.  
That is, we can see that $\rho_{2}$ is a symplectic structure on $M$.
%
Also, $B$-model
is constructed from a holomorphic three form $\rho_{3}$ on $6$ dimensional manifold $M$, which defines a complex structure on $M$.  
Let $\Sigma_3$ be a world volume in three dimensions.
%
The $2$-brane action of $B$ model is 
\begin{eqnarray}
S = \int_{\Sigma_3} \frac{1}{3!} \rho_{3, ijk}
d \phi^{i} d \phi^{j} d \phi^{k}.
\label{su3sdaction}
\end{eqnarray}
The first class constraint $\dM \rho_{3} = 0$ is 
the integrability condition of the holomorphic form.
This action is a dual membrane formulation of the $B$-model 
\cite{Bonelli:2005ti}.

 As the first example of our model, we consider $A$ model with $H \neq 0$.  We choose $n=2$.  
Our action (\ref{ggaction}) is 
\begin{eqnarray}
S = \int_{\Sigma_2}
e^{Y_2} e^{\frac{1}{2} b} \rho
= \int_{\Sigma_2}
\left[Y_2 \rho_{0}  +
\frac{1}{2!} (\rho_{2, ij}
+ b_{ij} \rho_{0}) d \phi^{i} d \phi^{j}
\right].
\label{su3eaction}
\end{eqnarray}
Here we will check that $\rho$ is a generalized structure (\ref{ggequation}).  
The consistency conditions (\ref{generalizedintegrablecomponent}) for $\rho \in \Lambda^{ev} (M)$ are
\begin{eqnarray}
&& \dM \rho_{0} = 0, 
\nonumber \\
&& \dM \rho_2 + H \wedge \rho_0 = 0.
\label{evensu31}
\end{eqnarray}
~From these two conditions and $\dM H = 0$, 
we obtain the equation $\dM (H \wedge \rho_2 ) = 0$.  
Therefore, from the Poicar\'e Lemma,
at least locally $\rho_4$ exists, s.t.  
\begin{eqnarray}
\dM \rho_4 + H \wedge \rho_2 = 0.
\label{evensu32}
\end{eqnarray}
Also, the equation 
\begin{eqnarray}
\dM \rho_6 + H \wedge \rho_4 = 0,
\label{evensu33}
\end{eqnarray}
is trivially satisfied in $6$ dimensions.
Therefore, A-type conditions (\ref{evensu31}), (\ref{evensu32}) and (\ref{evensu33}) for 
a generalized $SU(3)$ structure are satisfied on the $6$-manifold.

%
Second, we consider $B$ model with $H \neq 0$.  We choose $n=3$ in (\ref{ggaction}).  
%
The action is 
\begin{eqnarray}
S = \int_{\Sigma_3}
e^{Y_2} e^{\frac{1}{2} b} \rho
= \int_{\Sigma_3}
\left[Y_2 \rho_{1,i} d \phi^{i} +
\frac{1}{3!} (\rho_{3, ijk}
+ b_{[ij} \rho_{1,k]}) d \phi^{i} d \phi^{j} d \phi^{k}
\right].
\label{su3oaction}
\end{eqnarray}
The first class constraints (\ref{generalizedintegrablecomponent}) for $\rho \in \Lambda^{od} (M)$ are 
\begin{eqnarray}
&& \dM \rho_{1} = 0, 
\nonumber \\
&& \dM \rho_3 + H \wedge \rho_1 = 0.
\label{oddsu31}
\end{eqnarray}
Then since $\dM (H \wedge \rho_3 ) = 0$, at least locally $\rho_5$ exists s.t.
\begin{eqnarray}
\dM \rho_5 + H \wedge \rho_3 = 0.
\label{oddsu32}
\end{eqnarray}
The conditions (\ref{oddsu31}) and (\ref{oddsu32}) are B-type conditions for the generalized $SU(3)$ structure on $6$-manifold.

 Also, we have other realizations of generalized structures on $6$-manifold.
In the cases $n=4$ and $n=6$, we can similarly derive A-type generalized $SU(3)$ conditions on $M$.  
In the cases $n=5$, we can derive B-type generalized $SU(3)$ conditions on $M$.

\subsection{Topological Membranes on Generalized $G_2$-Manifolds}
\noindent
Next we consider the case of topological membrane theory with a generalized $G_2$ structure on a $7$ dimensional manifold $M$.

 First, we consider a $G_2$ structure on the $7$-manifold $M$, namely $H=0$.  
A $G_2$ structure is defined from a $3$-form $\rho_3$, which satisfies $d_M \rho_3 =0$ and $d_M * \rho_3 =0$.  
A $G_2$ structure is defined as the subgroup of $GL(7)$ to 
preserve $\rho_3$.
A manifold with a $G_2$ structure is called a $G_2$ manifold
\cite{Joyce}.  The worldvolume action of the topological M theory, which has a $G_2$ structure, was proposed as 
a topological $2$-brane \cite{Bonelli:2005ti}.  The action is 
%
\begin{eqnarray}
S = \int_{\Sigma_3}
\frac{1}{3!} \rho_{3, ijk}
d \phi^{i} d \phi^{j} d \phi^{k},
\label{g2action}
\end{eqnarray}
which is called self dual membrane \cite{Biran:1987ae}\cite{Grabowski:1989cx}.  
If a target space $M$ is a $G_2$ manifold, 
$d_M \rho_3 =0$ can be derived from the action (\ref{g2action}) as 
the first class constraint.

%
We consider the $n=3$ case in the action (\ref{ggaction}):
\begin{eqnarray}
S = \int_{\Sigma_3}
e^{Y_2} e^{\frac{1}{2} b} \rho
= \int_{\Sigma_3}
\left[Y_2 \rho_{1,i} d \phi^{i} +
\frac{1}{3!} (\rho_{3, ijk}
+ b_{[ij} \rho_{1,k]}) d \phi^{i} d \phi^{j} d \phi^{k}
\right].
\label{g3oaction}
\end{eqnarray}
This action gives B-type generalized $G_2$ conditions.  
In fact, the consistency conditions (\ref{generalizedintegrablecomponent}) are
\begin{eqnarray}
&& \dM \rho_{1} = 0, 
\nonumber \\
&& \dM \rho_3 + H \wedge \rho_1 = 0.
\label{oddg21}
\end{eqnarray}
~From these two equations, 
we obtain $\dM (H \wedge \rho_3 ) = 0$.  
Using the Poincar\'e Lemma, at least locally $\rho_5$ exists 
\begin{eqnarray}
\dM \rho_5 + H \wedge \rho_3 = 0.
\label{oddg22}
\end{eqnarray}
Also, the equation 
\begin{eqnarray}
\dM \rho_7 + H \wedge \rho_5 = 0,
\label{oddg23}
\end{eqnarray}
is trivially satisfied in $7$ dimensions.
Therefore we obtain (\ref{oddg21}), (\ref{oddg22}) and (\ref{oddg23}) as the B-type conditions for the generalized $G_2$ structure.

Next we consider a model with an A-type condition 
for a generalized $G_2$ structure.  For $\hat{\rho} \in \Lambda^{ev}(M)$, A-type condition is 
$(\dM + H) \wedge \hat{\rho} = 0$ \cite{Witt:2004vr}.  
%
%
Here, we use only the fact that 
$\hat{\rho}$ satisfies the condition
$(\dM + H) \wedge \hat{\rho} =0$.  The model can be constructed as follows.  
We choose a $n=4$ world volume $\Sigma_4$ in (\ref{ggaction}).
Here $n=4$ means to be dual to $n=3$ in $7$ dimensions. 
The action (\ref{ggaction}) is 
\begin{eqnarray}
S &=& \int_{\Sigma_4}
e^{Y_2} e^{\frac{1}{2} b} \hat{\rho}
\nonumber \\
&=& \int_{\Sigma_4}
\Bigl[
Y_2^2 \hat{\rho}_{0}
+ \frac{1}{2!} Y_2  (\hat{\rho}_{2, ij} + b_{ij} \hat{\rho}_{0})
d \phi^{i} d \phi^{j} 
\nonumber \\
&&
+ \frac{1}{4!} (\hat{\rho}_{4, ijkl} + b_{[ij} \hat{\rho}_{2, kl]})
d \phi^{i} d \phi^{j} d \phi^{k} d \phi^{l}
\Bigr].
\label{g2eaction}
\end{eqnarray}
The first class constraints for $\hat{\rho} \in \Lambda^{ev} (M)$ in 
(\ref{generalizedintegrablecomponent}) are 
\begin{eqnarray}
&& \dM \hat{\rho}_{0} = 0, 
\nonumber \\
&& \dM \hat{\rho}_2 + H \wedge \hat{\rho}_0 = 0,
\nonumber \\
&& \dM \hat{\rho}_4 + H \wedge \hat{\rho}_2 = 0.
\label{eveng21}
\end{eqnarray}
Since $\dM (H \wedge \hat{\rho}_4 ) = 0$ in $7$ dimensions,
we can locally derive the equation 
\begin{eqnarray}
\dM \hat{\rho}_6 + H \wedge \hat{\rho}_4 = 0.
\label{eveng22}
\end{eqnarray}
The conditions (\ref{eveng21}) and (\ref{eveng22}) 
are the A-type conditions for the generalized $G_2$ structure.
If $(\dM + H) \wedge \hat{\rho} =0$, the action is consistent.

Also, we have other realizations of generalized $G_2$ structures.
If we choose $n=5$ and $n=7$ in the action (\ref{ggaction}),
we can also derive B-type generalized $G_2$ conditions on $M$.
If we consider $n=2$ and $n=6$ in the action (\ref{ggaction}),
we can also derive A-type generalized $G_2$ conditions on $M$.

\subsection{Topological Membranes on Generalized $Spin(7)$-Manifolds}
\noindent
Let $M$ be a $8$ dimensional manifold.
A $Spin(7)$ structure 
is defined by a $Spin(7)$ $4$-form $\rho_4$,
which satisfies $d_M \rho_4 =0$ and selfdual condition $\rho_4 = *\rho_4$.
A $Spin(7)$ structure is defined as the subgroup of $GL(8)$ to 
preserve the $Spin(7)$ form $\rho_4$ \cite{Joyce}.  We can define a consistent topological $3$-brane model 
in a $4$ dimensional worldvolume. 
The action is
\begin{eqnarray}
S = \int_{\Sigma_4}
\frac{1}{4!} \rho_{4, ijkl}
d \phi^{i} d \phi^{j} d \phi^{k} d \phi^{l},
\label{spin7action}
\end{eqnarray}
where $\rho_4$ is a $Spin(7)$ form.
The first class constraints are equal to the $Spin(7)$ condition $\dM \rho_4 = 0$.
This action is considered as a worldvolume description of topological $F$ theory.

Topological membrane action on a generalized is constructed as follows.  
There are two types of generalized $Spin(7)$ structures, namely A-type and B-type.  
First we consider an A-type $Spin(7)$ structure,
which is defined by the sum of even forms $\rho \in \Lambda^{ev} (M)$.
By setting $n=4$ in (\ref{ggaction}), topological $3$-brane action is 
\begin{eqnarray}
S 
&=& \int_{\Sigma_4}
e^{Y_2} e^{\frac{1}{2} b} \rho
\nonumber \\
&=& \int_{\Sigma_4}
\Bigl[
Y_2 Y_2 \rho_{0}
+ \frac{1}{2!} Y_2  (\rho_{2, ij} + b_{ij} \rho_{0})
d \phi^{i} d \phi^{j}
\nonumber \\
&&
+ \frac{1}{4!} (\rho_{4, ijkl} + b_{[ij} \rho_{2, kl]})
d \phi^{i} d \phi^{j} d \phi^{k} d \phi^{l},
\Bigr].
\label{evenspin7}
\end{eqnarray}
The first class constraints are 
\begin{eqnarray}
&& \dM \rho_{0} = 0, 
\nonumber \\
&& \dM \rho_2 + H \wedge \rho_0 = 0,
\nonumber \\
&& \dM \rho_4 + H \wedge \rho_2 = 0.
\label{evenspin71}
\end{eqnarray}
~From these conditions, we can locally obtain the equations:
\begin{eqnarray}
&& \dM \rho_6 + H \wedge \rho_4 = 0,
\nonumber \\
&& \dM \rho_8 + H \wedge \rho_6 = 0.
\label{evenspin72}
\end{eqnarray}
The conditions (\ref{evenspin71}) and (\ref{evenspin72}) give an A-type generalized $Spin(7)$ structure.

Next we consider a B-type generalized
$Spin(7)$ structure.
In the previous subsection, 
a generalized $G_2$ structure is realized 
by use of topological $2$-brane and topological $3$-brane in the action (\ref{ggaction}).  
As we will discuss later, we can assume that that the method of dimensional reduction from a topological 
$F$ theory in $8$ dimensions to a topological $M$ theory in 
$7$ dimensions suggests two type of the $Spin(7)$ structures using 
a topological $3$ brane and a topological $4$ brane.
Therefore, we choose $n=5$ for 
a B-type generalized $Spin(7)$ structure
in (\ref{ggaction}).  The action is
\begin{eqnarray}
S &=& \int_{\Sigma_5}
e^{Y_2} e^{\frac{1}{2} b} \rho
\nonumber \\
&=& \int_{\Sigma_5}
\Bigl[Y_2 Y_2  \rho_{1,i} d \phi^{i} 
+ \frac{1}{3!} Y_2  (\rho_{3, ijk}
+ b_{[ij} \rho_{1,k]}) d \phi^{i} d \phi^{j} d \phi^{k}
\nonumber \\
&& + \frac{1}{5!} (\rho_{5, ijklm}
+ b_{[ij} \rho_{3, klm]} 
+ b_{[ij} b_{kl} \rho_{1,m]}) 
d \phi^{i} d \phi^{j} d \phi^{k}
\Bigr].
\label{oddspin7}
\end{eqnarray}
The first class constraints for $\rho \in \Lambda^{od} (M)$ are 
\begin{eqnarray}
&& \dM \rho_{1} = 0, 
\nonumber \\
&& \dM \rho_3 + H \wedge \rho_1 = 0,
\nonumber \\
&& \dM \rho_5 + H \wedge \rho_3 = 0.
\label{oddspin71}
\end{eqnarray}
~From these conditions, we obtain the equation
\label{oddspin72}
\begin{eqnarray}
&& \dM \rho_7 + H \wedge \rho_5 = 0.
\end{eqnarray}
The conditions (\ref{oddspin71}) and (\ref{oddspin72}) give a B-type generalized $Spin(7)$ structure.

Also, we have other realizations of generalized $Spin(7)$ structures.
When we choose $n=2$, $n=6$ or $n=8$ in the action (\ref{ggaction}),
we can derive A-type generalized $Spin(7)$ conditions on $M$.
When we choose $n=3$ or $n=7$ in the action (\ref{ggaction}),
we can derive B-type generalized $Spin(7)$ conditions on $M$.

\section{Dimensional Reduction}
\noindent
Here we consider dimensional reduction of our models from 
compactification of a target space $M$. Our generalized topological
membranes models are related to each other.  
There are two types of reductions.  
One is a wrapping membrane, and the other is a longitudinal membrane.  
These reductions from higher dimensional topological generalized membranes give lower dimensional A-type and B-type topological generalized membranes.  
Here we consider one-dimensional compactification 
$M_d = M_{d-1} \times S^1$,
where $M_d$ is a $d$ dimensional manifold
and $M_{d-1}$ is a $d-1$ dimensional manifold.

\subsection{Wrapping Membranes}
\noindent
First we consider topological $n-1$-brane wrapping to 
$S^1$.
Let $(\sigma^1, \cdots, \sigma^{n})$ be a local coordinate on a 
$n-1$-brane $\Sigma_n$.
$\sigma^n$ is chosen as a local coordinate of wrapping direction $S^1$. 
We use indices $i, j, k = 1, \cdots, d$ and 
$r, s = 1, \cdots, d-1$.
Local coordinates of the wrapping membrane are
\begin{eqnarray}
\phi^{i} = \left\{
\begin{array}{ll}
\phi^{r}(\sigma^1, \cdots, \sigma^{n-1}) 
& \mbox{if $r = 1, \cdots, d-1$} \\
\sigma^{n} & \mbox{if $r = d$}
\end{array}
\right.
\end{eqnarray}
and 
\begin{eqnarray}
&& b_{ij} = \left\{
\begin{array}{ll}
b_{rs}(\sigma^1, \cdots, \sigma^{n-1}) 
& \mbox{if $r, s = 1, \cdots, d-1$} \\
0 & \mbox{otherwise}
\end{array}
\right.
\nonumber \\
&&
Y_2 = Y_2(\sigma^1, \cdots, \sigma^{n-1}).
\end{eqnarray}
Then the action (\ref{ggaction}) is reduced to 
\begin{eqnarray}
S = 
\int_{\Sigma_n}
e^{Y_2} e^{\frac{1}{2} b} \rho
&=& \sum_{p=0}^{d} \frac{1}{p!} \int_{\Sigma_n}
e^{Y_2} e^{\frac{1}{2} b} 
 \rho_{p, i_1 \cdots i_p}(\phi)
d \phi^{i_1} \cdots d \phi^{i_p},
\nonumber \\
&=& \sum_{p=1}^{d} \frac{1}{(p-1)!} \left[
\int_{\Sigma_{n-1}}
e^{Y_2} e^{\frac{1}{2} b} 
 \rho_{p, i_1 \cdots i_{p-1} d}(\phi)
d \phi^{i_1} \cdots d \phi^{i_{p-1}} \right]
\int_{S^1} d \sigma ^n,
\end{eqnarray}
where $\Sigma_{n} = \Sigma_{n-1} \times S^1$.
We can see that a generalized structure on $M_d$ reduces to a generalized 
structure on $M_{d-1}$.  Also, this reduction changes a A-types generalized structure and a B-types generalized structure.  
For examples, a generalized A-type (B-type) $G_2$ membrane is reduced to 
a generalized B-type (A-type) $SU(3)$ string(membrane).  
A generalized A-type (B-type) $Spin(7)$ membrane is reduced to 
a generalized B-type (A-type) $G_2$ membrane.

\subsection{Longitudinal Membranes}
\noindent
Next we consider topological membrane longitudinal to a compactified direction $S^1$.
Local coordinates of the longitudinal membrane are 
\begin{eqnarray}
\phi^{i} = \left\{
\begin{array}{ll}
\phi^{r}(\sigma^1, \cdots, \sigma^{n}) 
& \mbox{if $r = 1, \cdots, d-1$} \\
0 & \mbox{if $r = d$}
\end{array}
\right.
\end{eqnarray}
and 
\begin{eqnarray}
&& b_{ij} = \left\{
\begin{array}{ll}
b_{rs}(\sigma^1, \cdots, \sigma^{n}) 
& \mbox{if $r, s = 1, \cdots, d-1$} \\
0 & \mbox{otherwise}
\end{array}
\right.
\nonumber \\
&&
Y_2 = Y_2(\sigma^1, \cdots, \sigma^{n}).
\end{eqnarray}
Then the action (\ref{ggaction}) is reduced to
\begin{eqnarray}
S = 
\int_{\Sigma_n}
e^{Y_2} e^{\frac{1}{2} b} \rho
&=& \sum_{p=0}^{d} \frac{1}{p!} \int_{\Sigma_n}
e^{Y_2} e^{\frac{1}{2} b} 
 \rho_{p, i_1 \cdots i_p}(\phi)
d \phi^{i_1} \cdots d \phi^{i_p},
\nonumber \\
&=& \sum_{p=0}^{d} \frac{1}{p!}
\int_{\Sigma_{n}}
e^{Y_2} e^{\frac{1}{2} b_{rs} d \phi^{r} d \phi^{s}} 
 \rho_{p, r_1 \cdots r_p}(\phi)
d \phi^{r_1} \cdots d \phi^{r_p} ,
\end{eqnarray}
where $r, s = 1, \cdots, d-1$.  
A generalized structure on $M_d$ is reduced to a generalized 
structure on $M_{d-1}$.  This reduction does not change an A-types and a B-types.  
For examples, a generalized A-type (B-type) $G_2$ membrane is reduced to 
a generalized A-type (B-type) $SU(3)$ membrane.
A generalized A-type (B-type) $Spin(7)$ membrane is reduced to 
a generalized A-type (B-type) $G_2$ membrane.

\section{AKSZ Formulation of Batalin-Vilkovisky Formalism}
\noindent
In order to construct and analyze models systematically, 
it is useful to use Batalin-Vilkovisky formalism.
In the case of topological field theories, 
the AKSZ formulation \cite{Alexandrov:1995kv} is a theoretical and
general setting of the Batalin-Vilkovisky formalism.

\subsection{Batalin-Vilkovisky Structures on Graded Vector Bundles}
\noindent
We explain a general setting of the AKSZ formulation
of the Batalin-Vilkovisky formalism for a general graded bundle
\cite{Ikeda:2006wd}.

Let $M$ be a smooth manifold in $d$ dimensions.
We define a {\it supermanifold} $\Pi T^* M$.
Mathematically, $\Pi T^* M$, whose bosonic part is $M$, is defined as 
a cotangent bundle with reversed parity of the fiber.
That is, a base manifold $M$ has Grassman even coordinates and 
the fiber of $\Pi T^* M$ has Grassman odd coordinates.
We introduce a grading called {\it total degrees}.
The coordinates of the base manifold have grade zero and 
the coordinates of the fiber have grade one.
Similarly, we can define $\Pi T M$ for a tangent bundle $T M$.

We must consider more general assignments 
for the degree of the fibers of $T^* M$ or $T M$.
For an 
integer $p$, we define $T^* [p] M$, which is called 
a graded cotangent bundle.
$T^* [p] M$ is a cotangent bundle, whose fiber has the degree $p$.
The coordinates of the bass manifold have the total degree zero and 
the coordinates of the fiber have the total degree $p$.
If $p$ is odd, the fiber is Grassman odd, and if $p$ is even, 
the fiber is Grassman even. 
We define a graded tangent bundle $T [p] M$ in the same way.
%
If we consider 
a general vector bundle $E$, a graded vector bundle $E[p]$ is defined 
in the similar way.
$E[p]$ is a vector bundle whose fiber has a shifted degree by $p$.  
Note that only the degree of fiber is shifted, and 
the degree of base space is not shifted. 

We consider a Poisson manifold $N$ with a Poisson bracket $\{*,*\}$.
If we shift the total degree, 
we can construct a graded manifold $\tilde{N}$ from $N$.  
Then the Poisson structure $\{*,*\}$ shifts to
a graded Poisson structure by grading of $\tilde{N}$.
The graded Poisson bracket is called an {\it antibracket} 
and denoted by $\sbv{*}{*}$.
$\sbv{*}{*}$ is graded symmetric and satisfies the graded Leibniz
rule and the graded Jacobi identity with respect to grading of the manifold. 
%
The antibracket $\sbv{*}{*}$ with 
 the total degree $- n + 1$ satisfies the following identities:
\begin{eqnarray}
&& \sbv{F}{G} = -(-1)^{(|F| + 1 - n)(|G| + 1 - n)} \sbv{G}{F},
\nonumber \\
&& \sbv{F}{G  H} = \sbv{F}{G} H
+ (-1)^{(|F| + 1 - n)|G|} G \sbv{F}{H},
\nonumber \\
&& \sbv{F  G}{H} = F \sbv{G}{H}
+ (-1)^{|G|(|H| + 1 - n)} \sbv{F}{H} G,
\nonumber \\
&& (-1)^{(|F| + 1 - n)(|H| + 1 - n)} \sbv{F}{\sbv{G}{H}}
+ {\rm cyclic \ permutations} = 0,
\label{BVidentity}
\end{eqnarray}
where $F, G$ and $H$ are functions on $\tilde{N}$, and
$|F|, |G|$ and $|H|$ are total degrees of the functions, respectively.
The graded Poisson structure is also called {\it P-structure}.

Typical examples of Poisson manifold $N$ are a cotangent bundle $T^* M$ 
and a vector bundle $E \oplus E^*$.  
%
First we consider a cotangent bundle $T^* M$.
Since $T^* M$ has a symplectic structure,
we can define a Poisson bracket induced from the natural 
symplectic structure.
If we take local coordinates $\phi^i$ on M and local coordinates 
$B_i$ of the fiber, we can define a Poisson bracket as follows:
%
\begin{eqnarray}
\{ F, G \} \equiv 
F \frac{\rd}{\partial \phi^i} 
\frac{\ld }{\partial B_i}  G
- 
F \frac{\rd }{\partial B_i} 
\frac{\ld }{\partial \phi^i} G,
\label{ab0ike}
\end{eqnarray}
where $F$ and $G$ are functions on $T^* M$, and
${\rd}/{\partial \varphi}$ and ${\ld}/{\partial
\varphi}$ are the right and left differentiations
with respect to $\varphi$, respectively.
%
Here we shift the degree of fiber by $p$, 
i.e. the space $T^* [p]M$.
Then the Poisson structure shifts to a graded Poisson structure.
The corresponding graded Poisson bracket is called {\it antibracket}, 
$\sbv{*}{*}$.
%
Let $\bphi^i$ be local coordinates of $M$
 and $\bb_{n-1,i}$ a basis of the fiber of $T^* [p] M$.
The antibracket $\sbv{*}{*}$ 
on a cotangent bundle $T^* [p] M$ is expressed as:
%
\begin{eqnarray}
\sbv{F}{G} \equiv 
F \frac{\rd}{\partial \bphi^i} 
\frac{\ld }{\partial \bb_{p,i}}  G
- 
F \frac{\rd }{\partial \bb_{p,i}} 
\frac{\ld }{\partial \bphi^i} G.
\label{Poisson2ike}
\end{eqnarray}
The total degree of the antibracket $\sbv{*}{*}$ is $-p$.
This antibracket satisfies the property (\ref{BVidentity})
for $-p = -n +1$.


Next, we consider a vector bundle $E \oplus E^*$. 
There is a natural Poisson structure on the fiber of $E \oplus E^*$ induced 
~from a paring of $E$ and $E^*$.
If we take local coordinates $A^a$ on the fiber of $E$ and 
$B_a$ on the fiber of $E^*$, we can define
\begin{eqnarray}
\{ F, G \} \equiv 
F \frac{\rd}{\partial A^a} 
\frac{\ld }{\partial B_a}  G
- 
F \frac{\rd }{\partial B_a} 
\frac{\ld }{\partial A^a} G,
\label{abpike}
\end{eqnarray}
where $F$ and $G$ are functions on $E \oplus E^*$.
We shift the degrees of fibers of $E$ and $E^*$ like $E [p] \oplus E^*[q]$, where
$p$ and $q$ are positive integers.
The Poisson structure changes to a graded Poisson structure
$\sbv{*}{*}$.
Let $\ba_p{}^{a}$ be a basis of the fiber of 
$E[p]$ and 
$\bb_{q,a}$ a basis of the fiber of 
$E^*[q]$.
The antibracket is represented as
\begin{eqnarray}
\sbv{F}{G} \equiv 
F  \frac{\rd}{\partial \ba_p{}^{a}} 
\frac{\ld }{\partial \bb_{q,a}}  G
- (-1)^{p q}
F \frac{\rd }{\partial \bb_{q,a}} 
\frac{\ld }{\partial \ba_p{}^{a}}  G.
\label{Poisson4ike}
\end{eqnarray}
The total degree of the antibracket $\sbv{*}{*}$ is $-p-q$.
This antibracket satisfies the property (\ref{BVidentity})
for $-p-q = -n +1$.


We define a {\it Q-structure}.
A {\it Q-structure} is 
a function $S$ on a graded manifold $\tilde{N}$
which satisfies the classical master equation 
$\sbv{S}{S} = 0$.
$S$ is called a {\it Batalin-Vilkovisky action}.  
We require that $S$ satisfy the compatibility condition
\begin{eqnarray}
S \sbv{F}{G} = \sbv{S F}{G} + (-1)^{|F| +1} \sbv{F}{SG},
\end{eqnarray}
where $F$ and $G$ are arbitrary functions, and 
$|F|$ is the total degree of $F$.
$\sbv{S}{F} = \brs F$ generates an infinitesimal transformation.
We call this a {\it BRST transformation}, which
coincides with the gauge transformation of the theory.

The AKSZ formulation of the Batalin-Vilkovisky formalism is defined as 
a {\it P-structure} and a {\it Q-structure} on a {\it graded manifold}.


\subsection{Batalin-Vilkovisky Structures of Topological Sigma Models}
\noindent
%
%
%
In this subsection, we explain Batalin-Vilkovisky structures 
of topological sigma models.
Let $X$ be a base manifold in $n$ dimensions, with or without boundary, 
and $M$ be a target manifold in $d$ dimensions.
We denote $\phi$ a smooth map from $X$ to $M$.
%

We consider a {\it supermanifold} $\Pi T X$, whose bosonic part is $X$.
$\Pi T X$ is defined as 
a tangent bundle with reversed parity of the fiber.
We extend a smooth function $\phi$ to a function on the 
supermanifold $\bphi:\Pi T X \rightarrow M$.
$\bphi$ is an element of $\Pi T^* X \otimes M$.
The {\it total degree} defined in the previous section is a grading 
with respect to $M$.
We introduce a new non-negative integer grading on $\Pi T^* X$.
A coordinate on a base manifold has zero 
and a coordinate on the fiber has one.
This grading is called {\it form degrees}.
We denote ${\rm deg} F$ the form degree of the function $F$.
${\rm gh} F = |F| - {\rm \deg} F$ is called {\it ghost number}.

First we consider a {\it P-structure} on $T^* [p] M$.
We take $p=n-1$ to construct
a Batalin-Vilkovisky structure in a topological sigma model
on a general $n$ dimensional worldvolume.
We consider $T^* [n-1] M$ for an $n$-dimensional base manifold $X$.
Let $\bphi^i$ be local coordinates of $\Pi T^* X \otimes M$, 
where $i, j, k, \cdots$ are indices of the local coordinates on $M$.
Let $\bb_{n-1,i}$ be a basis 
of sections of $\Pi T^* X \otimes \bphi^*(T^* [n-1] M)$.
As we discussed in the previous subsection, 
we can define an {\it antibracket} $\sbv{*}{*}$ 
on a cotangent bundle $T^* [n-1] M$ as
%
\begin{eqnarray}
\sbv{F}{G} \equiv 
F \frac{\rd}{\partial \bphi^i} 
\frac{\ld }{\partial \bb_{n-1,i}}  G
- 
F \frac{\rd }{\partial \bb_{n-1,i}} 
\frac{\ld }{\partial \bphi^i} G,
\label{cotangentPike}
\end{eqnarray}
where 
$F$ and $G$ are functions of $\bphi^i$ and $\bb_{n-1,i}$.
The total degree of the antibracket is $-n+1$.
If $F$ and $G$ are functionals of 
$\bphi^i$ and $\bb_{n-1,i}$, 
we understand an antibracket is defined as 
\begin{eqnarray}
\sbv{F}{G} \equiv 
\int_{X} 
F \frac{\rd}{\partial \bphi^i} 
\frac{\ld }{\partial \bb_{n-1,i}}  G
- 
F \frac{\rd }{\partial \bb_{n-1,i}} 
\frac{\ld }{\partial \bphi^i} G,
\end{eqnarray}
where the integration over $X$ pick up only the 
$n$-form part of the integrand. 
Through this article, 
we always understand an antibracket on two functionals in a similar 
manner and abbreviate this notation.

Next we consider a {\it P-structure} on $E \oplus E^*$.
We assign the total degrees
$p$ and $q$ such that $p+q = n-1$.  
That is, 
we consider $E[p] \oplus E^*[n-p-1]$, 
where $-1 \leq p \leq n-1$.  
Then we can construct a topological sigma model.  
%

%
Let $\ba_p{}^{a_p}$ be a basis of sections of 
$\Pi T^* X \otimes \bphi^*(E[p])$ and 
$\bb_{n-p-1,a_p}$ a basis of the fiber of 
$\Pi T^* X \otimes \bphi^*(E^*[n-p-1])$.
~From (\ref{Poisson4ike}), we can define the antibracket 
as
\begin{eqnarray}
\sbv{F}{G} \equiv 
F  \frac{\rd}{\partial \ba_p{}^{a_p}} 
\frac{\ld }{\partial \bb_{n-p-1,a_p}}  G
- (-1)^{n p}
F \frac{\rd }{\partial \bb_{n-p-1,a_p}} 
\frac{\ld }{\partial \ba_p{}^{a_p}}  G.
\label{EEPike}
\end{eqnarray}


We need to consider various grading assignments for $E \oplus E^*$,
because each assignment 
induces different Batalin-Vilkovisky structures.
In order to consider all independent assignments, 
we define the following bundle.
Let $E_p$ be 
$n$
series of vector bundles, 
where 
$-1 \leq p \leq n-1$.  
We consider a direct sum  of each bundle
$E_p[p] \oplus E_p^*[n-p-1]$ :
\begin{eqnarray}
\sum_{p=-1}^{n-1}
E_p[p] \oplus E_p^*[n-p-1].
\label{totbundleike}
\end{eqnarray}
We define a {\it P-structure}
on the graded vector bundle
\begin{eqnarray}
\left(
\sum_{p=-1}^{n-1}
E_p[p] \oplus E_p^*[n-p-1] \right) \oplus T^*[n-1] M.
\label{totspaceike}
\end{eqnarray}
A local (Darboux) coordinate expression for
the antibracket $\sbv{\cdot}{\cdot}$ is a sum of 
(\ref{cotangentPike}) and (\ref{EEPike}):
\begin{eqnarray}
\sbv{F}{G} \equiv 
\sum_{p=-1}^{n-1}
F \frac{\rd}{\partial \ba_p{}^{a_p}} 
\frac{\ld }{\partial \bb_{n-p-1 \ a_p}} G
- (-1)^{n p}
F \frac{\rd }{\partial \bb_{n-p-1 \ a_p}} 
\frac{\ld }{\partial \ba_p{}^{a_p}} G.
\label{bfantibracketike}
\end{eqnarray}
where 
$p=0$ component is the antibracket (\ref{cotangentPike}) on 
the graded cotangent bundle $T^* [n-1] M$,
and $\ba_0{}^{a_0} = \bphi^i$.
Note that all terms of the antibracket have 
the total degree $-n+1$, and 
we can confirm that 
the antibracket (\ref{bfantibracketike}) satisfies the identity
(\ref{BVidentity}).

\section{Batalin-Vilkovisky Formulation of Topological Membranes
with Generalized Structures}
\noindent
We consider the target graded bundle
$(T[1]M \oplus T^*[n-2]M) \oplus (M \times \bR[n-3] \oplus M \times \bR[2])
\oplus T^*[n-1]M$.
i.e. 
We take $E_1 = TM$, $E_{n-3} = M \times \bR$
and $E_p =0$ for the other $p$ in (\ref{totspaceike}).
We introduce the following superfields 
which are sections on the above total bundle:
\hfil\break
$\bphi:\Pi TX \rightarrow M$, 
$\bb_{n-1,i} \in \Gamma(\Pi T^* X \otimes \bphi^* (T^*[n-1] M))$, 
$\ba_1{}^i \in \Gamma(\Pi T^* X \otimes \bphi^* (T[1] M))$, 
$\bb_{n-2,i} \in \Gamma(\Pi T^* X \otimes \bphi^* (T^*[n-2] M))$, 
$\bz_{n-3} \in \Pi T^* X \otimes \bphi^* (M \times \bR[n-3])$
and 
$\by_{2} \in \Gamma(\Pi T^* X \otimes \bphi^* (M \times \bR[2]))$.

The antibracket is derived from (\ref{bfantibracketike}) as:
\begin{eqnarray}
\sbv{F}{G} &\equiv&
F  \frac{\rd}{\partial \bphi^{i}} 
\frac{\ld }{\partial \bb_{n-1,i}}  G
- 
F \frac{\rd }{\partial \bb_{n-1,i}} 
\frac{\ld }{\partial \bphi^{i}}  G
\nonumber \\ &&
+
F  \frac{\rd}{\partial \ba_1{}^{i}} 
\frac{\ld }{\partial \bb_{n-2,i}}  G
- (-1)^n
F \frac{\rd }{\partial \bb_{n-2,i}} 
\frac{\ld }{\partial \ba_1{}^{i}}  G
\nonumber \\ &&
+ F  \frac{\rd}{\partial \bz_{n-3}} 
\frac{\ld }{\partial \by_{2}}  G
- 
F \frac{\rd }{\partial \by_{2}} 
\frac{\ld }{\partial \bz_{n-3}}  G.  
\end{eqnarray}
This defines a total degree $-n+1$ {\it P-structure}.

We remember the $1$-st order formalism action (\ref{ndaction}).
By the superfield extension of the action (\ref{ndaction}),
we can propose the following $1$-st order BV action 
\begin{eqnarray}
S &=& \int_{\Sigma_n}
(-1)^{n} \bb_{n-1,i} d \bphi{}^i + (-1)^{n-1} \bb_{n-2,i} d \ba_1{}^i
- \by_{2} d \bz_{n-3}
+ (-1)^{n-1} \bb_{n-1,i} \ba_1{}^{i}
\nonumber \\
&& 
+ \exp{(\by_2)} \exp{\left( \frac{1}{2} b_{ij} \ba_1{}^i \ba_1{}^j \right)}
\biggl(
\sum_{p=0}^{d}
\frac{1}{p!} \rho_{p, i_1 \cdots i_p}(\bphi)
\ba_1{}^{i_1} \cdots \ba_1{}^{i_p}
\biggr),
\label{bvndaction}
\end{eqnarray}
where $b_{ij}(\bphi)$ and $\rho_{p, i_1 \cdots i_p}(\bphi)$ are 
functions of $\bphi$.
Then the classical master equation is
\begin{eqnarray}
\sbv{S}{S} = 0.
\end{eqnarray}
This provides the Batalin-Vilkovisky structure in this model.
This master equation is 
%
\begin{eqnarray}
&& \frac{\ld }{\partial \bphi^j} \rho_{p+2, i_1 \cdots i_{p+2}}(\bphi)
+ \frac{3!}{(p+2)(p+1)} 
H_{[ji_{p+1}i_{p+2}}(\bphi) \rho_{p, i_1 \cdots i_{p}]}(\bphi)
= 0.
\label{bvgeneralized}
\end{eqnarray}
This equation is nothing but the equation 
(\ref{generalizedintegrablecomponent}) 
for $b_{ij}(\bphi)$ and $\rho_{p, i_1 \cdots i_p}(\bphi)$.  
Therefore this topological sigma model (\ref{bvndaction})
defines the generalized geometry structure on the target space $M$
as the Batalin-Vilkovisky structure.  

\subsection{BV Actions with Generalized $SU(3)$ Structures}
\noindent
We consider a BV construction of the models with 
a generalized  $SU(3)$ structure, which we proposed in section 4.1.  
First we consider the case of $n=2$ in (\ref{bvndaction}).
$\Sigma_2$ 
is a two dimensional worldsheet.  This model is a  topological string.  
The action is 
\begin{eqnarray}
S &=& \int_{\Sigma_2}
\bb_{1,i} d \bphi{}^i - \bb_{0,i} d \ba_1{}^i
- \by_{2} d \bz_{-1}
- \bb_{1,i} \ba_1{}^{i}
\nonumber \\
&& 
+ \exp{(\by_2)} \exp{(\frac{1}{2} b_{ij} \ba_1{}^i \ba_1{}^j)}
\biggl(\rho_{0}(\bphi) 
+ \frac{1}{2!} \rho_{2, ij}(\bphi) \ba_1{}^i \ba_1{}^j
\biggr)
\nonumber \\
&=&
\int_{\Sigma_2}
\bb_{1,i} d \bphi{}^i - \bb_{0,i} d \ba_1{}^i
- \by_{2} d \bz_{-1}
- \bb_{1,i} \ba_1{}^{i}
\nonumber \\
&& 
+ \biggl( \rho_{0} \by_2 +
\frac{1}{2!} (\rho_{2, ij}
+ b_{ij} \rho_{0}) \ba_1{}^i \ba_1{}^j
\biggr).
\label{2dbvaction}
\end{eqnarray}
This action is a BV construction of the action (\ref{su3eaction}) 
and has an A-type condition for 
a generalized $SU(3)$ structure.

Also, this action is an extension of the $A$ model by use of the AKSZ superfield 
formulation.  
The $A$ model action based on a symplectic form $Q_{ij}$
is defined as
\begin{eqnarray}
S_{A} = \int_{\Sigma}
\bb_{1i} d \bphi^i - \bb_{0i} d \ba_1^i 
- \bb_{1i} \ba_1^i 
+ \frac{1}{2} Q_{ij}(\bphi) \ba_1^i \ba_1^j. 
\label{aqmodel}
\end{eqnarray}
We can see that the classical master equation $\sbv{S_{A}}{S_{A}} = 0$ is satisfied, if and only if 
the $2$-form 
$Q = Q_{ij} d \bphi^i d \bphi^j$ on the target space is 
a symplectic form $d_M Q = 0$, 
i.e.
$
\partial_k Q_{ij} + \partial_i Q_{jk} + \partial_j Q_{ki} = 0
$.  
%
When we set $b_{ij} = 0$,
$\by_{2}$ and $\bz_{-1}$ decouple from the other fields, and 
the remaining part of the action (\ref{2dbvaction}) 
coincides with $A$ model (\ref{aqmodel}).  
(\ref{2dbvaction}) is a $B$-field deformation of $A$ model.


Next we derive a B-type condition for a generalized $SU(3)$ 
structure.
Let a target space $M$ have even dimensions, especially 
$6$ dimensions.
If we set $n=3$ in (\ref{bvndaction}), 
we obtain the BV action of the action (\ref{su3oaction}) 
:
\begin{eqnarray}
S &=& \int_{\Sigma_3}
- \bb_{2,i} d \bphi{}^i + \bb_{1,i} d \ba_1{}^i
- \by_{2} d \bz_0
+ \bb_{2,i} \ba_1{}^{i}
\nonumber \\
&& 
+ \exp{(\by_2)} \exp{(\frac{1}{2} b_{ij} \ba_1{}^i \ba_1{}^j)}
\biggl(
\sum_{q=0}^{1}
\frac{1}{(2q+1)!} \rho_{2q+1, i_1 \cdots i_{2q+1}}(\bphi)
\ba_1{}^{i_1} \cdots \ba_1{}^{i_{2q+1}}
\biggr).
\label{3dbvactionodd}
\end{eqnarray}
This action is a B-type membrane action with a generalized 
$SU(3)$ structure.

\subsection{BV Actions for Generalized $G_2$ Structures}
\noindent
Let a target space $M$ have $7$ dimensions.
If we set $n=3$ in the action (\ref{bvndaction}),
we obtain the BV action 
for the action (\ref{g3oaction}), which has 
a B-type generalized $G_2$ structure:
\begin{eqnarray}
S &=& \int_{\Sigma_3}
- \bb_{2,i} d \bphi{}^i + \bb_{1,i} d \ba_1{}^i
- \by_{2} d \bz_0
+ \bb_{2,i} \ba_1{}^{i}
\nonumber \\
&& 
+ \exp{(\by_2)} \exp{(\frac{1}{2} b_{ij} \ba_1{}^i \ba_1{}^j)}
\biggl(
\sum_{q=0}^{1}
\frac{1}{(2q+1)!} \rho_{2q+1, i_1 \cdots i_{2q+1}}(\bphi)
\ba_1{}^{i_1} \cdots \ba_1{}^{i_{2q+1}}
\biggr).
\label{3dbvactioneven}
\end{eqnarray}
When we set $b_{ij} = 0$,
$\by_{2}$ and $\bz_{0}$ decouple from the other fields, and 
the remaining part of the action (\ref{3dbvactioneven}) is
\begin{eqnarray}
S = \int_{\Sigma_3}
- \bb_{2,i} d \bphi{}^i + \bb_{1,i} d \ba_1{}^i
+ \bb_{2,i} \ba_1{}^{i}
+ \frac{1}{3!} \rho_{3, ijk}(\bphi) \ba_1{}^i \ba_1{}^j \ba_1{}^k.
\label{g2bvmembraneaction}
\end{eqnarray}
The master equation $\sbv{S}{S} = 0$ is equivalent to
the $G_2$ condition $\dM \rho_3 =0$. 
The action (\ref{g2bvmembraneaction})
is a BV action for a $G_2$ topological membrane.  
The action (\ref{3dbvactioneven})
is a $B$-field deformation of $G_2$ topological membrane.

Also, if we consider worldvolume $\Sigma_n$ in even dimensions, 
we obtain an A-type generalized $G_2$ structure.
When we set $n=4$, 
we obtain the BV action for the action (\ref{g2eaction}):
\begin{eqnarray}
S &=& \int_{\Sigma_4}
\bb_{3,i} d \bphi{}^i - \bb_{2,i} d \ba_1{}^i
- \by_{2} d \bz_{1}
- \bb_{3,i} \ba_1{}^{i}
\nonumber \\
&& 
+ \exp{(\by_2)} \exp{( \frac{1}{2} b_{ij} \ba_1{}^i \ba_1{}^j)}
\biggl(
\sum_{q=0}^{2}
\frac{1}{(2q)!} \rho_{2q, i_1 \cdots i_{2q}}(\bphi)
\ba_1{}^{i_1} \cdots \ba_1{}^{i_{2q}}
\biggr).
\label{4dbvactionodd}
\end{eqnarray}
This action is a A-type membrane action with a generalized $G_2$ structure.  
\subsection{BV Actions for Generalized $Spin(7)$ Structures}
\noindent
Let a target space $M$ have $8$ dimensions.
If we set $n=4$ in the general action (\ref{bvndaction}), 
we obtain the BV action for the action
(\ref{evenspin7}):
\begin{eqnarray}
S &=& \int_{\Sigma_4}
\bb_{3,i} d \bphi{}^i - \bb_{2,i} d \ba_1{}^i
- \by_{2} d \bz_{1}
- \bb_{3,i} \ba_1{}^{i}
\nonumber \\
&& 
+ \exp{(\by_2)} \exp{\frac{1}{2} (b_{ij} \ba_1{}^i \ba_1{}^j)}
\biggl(
\sum_{q=0}^{2}
\frac{1}{(2q)!} \rho_{2q, i_1 \cdots i_{2q}}(\bphi)
\ba_1{}^{i_1} \cdots \ba_1{}^{i_{2q}}
\biggr).
\label{4dbvspin7actioneven}
\end{eqnarray}
This action is a BV action with an A-type generalized $Spin(7)$ 
structure.

When we set $b_{ij} = 0$,
$\by_{2}$ and $\bz_{1}$ decouple from the other fields, and 
the remaining part of the action (\ref{4dbvspin7actioneven}) is
\begin{eqnarray}
S = \int_{\Sigma_4}
\bb_{3,i} d \bphi{}^i - \bb_{2,i} d \ba_1{}^i
- \bb_{3,i} \ba_1{}^{i}
+ \frac{1}{4!} \rho_{4,ijkl}(\bphi) 
\ba_1{}^i \ba_1{}^j \ba_1{}^k \ba_1{}^l,
\label{4dbvspin7action}
\end{eqnarray}
This action is a BV action of $Spin(7)$ topological membrane.  
The master equation $\sbv{S}{S}=0$ of the action (\ref{4dbvspin7action}) is satisfied 
if $\rho_4$ is a $Spin(7)$-form.

Also, when we consider worldvolume $\Sigma_5$ and 
a target space $M$ with $8$ dimensions,
 the BV action for the action (\ref{oddspin7})is 
\begin{eqnarray}
S &=& \int_{\Sigma_5}
- \bb_{4,i} d \bphi{}^i + \bb_{3,i} d \ba_1{}^i
- \by_{2} d \bz_{2}
+ \bb_{4,i} \ba_1{}^{i}
\nonumber \\
&& 
+ \exp{(\by_2)} \exp{(\frac{1}{2} b_{ij} \ba_1{}^i \ba_1{}^j)}
\biggl(
\sum_{q=0}^{2}
\frac{1}{(2q+1)!} \rho_{2q+1, i_1 \cdots i_{2q+1}}(\bphi)
\ba_1{}^{i_1} \cdots \ba_1{}^{i_{2q+1}}
\biggr).
\label{4dbvspin7actionodd}
\end{eqnarray}
This action is a B-type membrane action with a generalized $Sipn(7)$ structure.  
\section{Conclusions and Discussions}
\noindent
We have proposed
topological string and topological membrane actions, 
which realize generalized geometries with nontrivial $3$-form flux $H$.  
The constraints of these actions are first class constraints if and 
only if the condition (\ref{ggequation}) is satisfied 
on the target manifolds $M$.
%
Especially, as examples, we have considered target manifolds $M$ with generalized $SU(3)$, $G_2$
and $Spin(7)$ structures.
Also, we have considered wrapping and longitudinal dimensional reductions.  
Then in our models, lower dimensional 
A-type and B-type generalized structures appear 
~from higher dimensional A-type and B-type 
generalized structures.  Finally, we have rewritten these actions by use of the
AKSZ formulation of Batalin-Vilkovisky formalism.


In this paper, we have considered only classical theory.
It would be an interesting work to quantize these theories as a quantization of generalized geometries.  
Also, it is known that some topological theories can be related 
to non-topological physical theories.  
For example, topological string theory on Calabi-Yau manifolds can be used to know physical information in type II superstring theory on Calabi-Yau backgrounds.  
%
%
%
%
%
%
In order to derive physical information in superstring theory on NS-NS $H$ flux background from topological string theory with $H$ flux,  
it would be a useful thing to find a relation with these.

%


\newcommand{\bibit}{\sl}


\vfill\eject
\end{document}